\documentclass[]{article}

\usepackage[utf8]{inputenc}
\usepackage{setspace}
\usepackage{epsfig}    
\usepackage{amsfonts}
\usepackage{amssymb}
\usepackage{amsmath}
\usepackage{subfigure}
\usepackage{comment}
\usepackage{multirow}
\usepackage{fullpage}
\usepackage{soul} 
\usepackage{graphicx}
\usepackage[dvipsnames]{xcolor}

\begin{document}

\title{AP-initiated Multi-User Transmissions in IEEE 802.11ax WLANs}
\author{Boris Bellalta$^a$ and Katarzyna Kosek-Szott$^b$\\
(a) Universitat Pompeu Fabra, Barcelona;\\
(b) AGH University of Science and Technology, Krakow, Poland.}

\date{}

\maketitle

\begin{abstract}
Next-generation IEEE 802.11ax Wireless Local Area Networks (WLANs) will make extensive use of multi-user communications in both downlink (DL) and uplink (UL) directions to achieve high and efficient spectrum utilization in scenarios with many user stations per access point. It will become possible with the support of multiuser (MU) multiple input, multiple output (MIMO) and orthogonal frequency division multiple access (OFDMA) transmissions. In this paper, we first overview the novel characteristics introduced by IEEE 802.11ax to implement AP-initiated OFDMA and multiuser MIMO (MU-MIMO) transmissions in both DL and UL directions. Namely, we describe the changes made at the physical layer and at the medium access control layer to support OFDMA, the use of \emph{trigger frames} to schedule uplink multi-user transmissions, and the new \emph{multiuser RTS/CTS mechanism} to protect large multi-user transmissions from collisions. Then, in order to study the achievable throughput of an IEEE 802.11ax WLAN, we use both mathematical analysis and simulations to numerically quantify the gains of MU transmissions and the impact of IEEE 802.11ax overheads on the WLAN saturation throughput. Results show the advantages of using MU transmissions in scenarios with many user stations. Additionally, we provide novel insights on the conditions in which IEEE 802.11ax WLANs are able to maximize their performance, such as the need to provide strict prioritization to AP-initiated MU transmissions to avoid collisions with transmissions from user stations.

\vspace{0.25cm}
\end{abstract}

\doublespacing
 
\section{Introduction}

% ---- Motivation

Wireless local area network (WLAN) technology is continuously evolving to keep the pace with an ever increasing number of users, traffic volume, new scenarios and use-cases. With the goal of offering a sustained multi-Gb/s aggregate throughput in scenarios with high density of access points (APs) and user stations (STAs), the IEEE 802.11 community created the IEEE 802.11ax Task Group (TGax) to develop a new set of physical (PHY) layer and medium access control (MAC) layer specifications. The new IEEE 802.11ax amendment, which is also called high efficiency (HE) WLANs, is currently available in a draft version \cite{IEEE80211axDraft}, and it is expected to be released by 2019.
 
The IEEE 802.11ax amendment is based on the IEEE 802.11ac-2013 \cite{IEEE80211ac}. It extends IEEE 802.11ac multi-user (MU) communication capabilities by including uplink multi-user multiple input, multiple output (UL MU-MIMO) and orthogonal frequency division Multiple access (OFDMA) techniques, among other improvements. MU transmissions allow to simultaneously serve multiple user stations at speeds compatible with their network interfaces (i.e., the rate at which the operating system/driver is able to deliver/read data to/from the IEEE 802.11 interface) and channel conditions. As a result, they allow achieving high aggregate throughput by reusing the same channel resources among multiple users, and thus minimizing packet and channel access protocol overheads. 

% ---- Minimal Related work

In addition to the TGax documents and the current version of the IEEE 802.11ax draft, in the literature there are several papers from the research community focusing on TGax developments, future WLAN scenarios and use cases, and the novel PHY/MAC enhancements proposed for IEEE 802.11ax WLANs \cite{khorov2015ieee,bellalta2016ieee,bellalta2016next,afaqui2016ieee,omarsurvey}. Additionally, the topic of IEEE 802.11 MU communications has been active in the last few years. A detailed survey of the most significant MU-MIMO-related solutions for WLANs is presented in \cite{liao2014mu}. A short survey of the OFDMA-related works is presented in \cite{li2015survey}. Other research works related to IEEE~802.11ax development focus on the performance of WLANs in dense scenarios. They include: the evaluation of the dynamic sensitivity control mechanism \cite{afaqui2015evaluation}, the use of basic service set (BSS) coloring \cite{selinis2016evaluation}, single-user (SU) performance \cite{sharon2017single}, scheduling \cite{karaca2016resource,sharon2017scheduling}, static and dynamic channel bonding \cite{bellalta2016interactions,faridi2015analysis,barrachina2018performance}, channel access configuration \cite{khorov2016several}, and more efficient but backward compatible alternatives to the legacy distributed coordination function~\cite{sanabria2014high,sanabria2016collision}.

% ---- Contribution

In this paper we focus on the analysis of MU transmissions in future IEEE 802.11ax WLANs with the goal of understanding the benefits and possible side effects, in terms of link layer overheads, that MU transmissions will bring to future WLANs. We consider both the OFDMA and MU-MIMO capabilities as defined in the current IEEE 802.11ax draft. Additionally, we describe the possible operation of future IEEE 802.11ax WLANs considering the case in which the AP is in charge of scheduling both downlink (DL) and UL MU transmissions. In such a case, the AP can properly apply quality of service (QoS) and load balancing policies to efficiently serve end users based on the knowledge of the network status. The downside of this approach, however, is that the AP needs to get from the user stations: (i) the channel state information (CSI) in both DL and UL directions, and (ii) the information about the availability of packets waiting for transmission, which may represent a significant overhead. To estimate the achievable IEEE 802.11ax saturation throughput, we extend Bianchi's IEEE 802.11 analytical model \cite{bianchi2000performance} with the new IEEE~802.11ax WLAN features when the proposed AP-initiated MU transmissions scheme is employed. By using the analytical model we obtain the DL and UL throughput, thus allowing us to explore different WLAN configurations to investigate how to improve IEEE 802.11ax WLAN performance. 

In detail, the contributions of this paper are the following:

\begin{enumerate}
	\item We introduce in a simple and clear way how IEEE 802.11ax MU transmissions work, including the new features such as the Multiuser Request to Send (MU-RTS) or the Multi-STA Block Acknowledgment (MS-BACK) procedures. 
	\item We adapt Bianchi's analytical model to evaluate the AP-initiated MU transmissions scheme proposed for IEEE 802.11ax WLANs, showing how to calculate the duration of successful transmissions and collisions, as well as including the overhead caused by the channel sounding procedure. 
	\item By showing the achievable throughput when only the AP is transmitting, we motivate the need of using large MU transmissions and packet aggregation to increase the efficiency of current WLANs.
	\item With the focus placed on the achievable performance of a single IEEE 802.11ax WLAN, we show the UL and DL saturation throughput as a function of the number of stations. We compare the obtained results with the IEEE 802.11ac WLAN saturation throughput. In order to provide a fair comparison, when the same parameters cannot be used in the two cases, we use the best configuration allowed by the considered amendment (i.e., channel width, number of spatial streams, maximum number of frames that can be aggregated in a single Aggregate MAC Protocol Data Unit (A-MPDU), etc.).
	\item We investigate the impact on the system performance of the new high-efficiency (HE) channel sounding protocol, which also benefits from the use of UL MU transmissions. We observe that the achieved gain compared with the case of transmitting the feedback from each user station sequentially is very high, resulting in very low channel sounding overheads even for a large number of user stations. 
	\item We investigate the impact of the channel width, maximum A-MPDU size, and the number of antennas at the AP on the IEEE 802.11ax WLAN performance. Among other results, we show that DL throughput can be severely harmed by allowing large A-MPDU transmissions.
	\item Finally, observing that AP-initiated MU UL transmissions may collide with UL SU transmissions, we propose to increase the CW$_{\min}$ parameter of user stations to $i$) avoid such collisions, $ii$) give more transmission opportunities to the AP. Results confirm that such an approach is able to increase the WLAN throughput since more MU transmissions can be scheduled.
\end{enumerate}

% ---- Paper Structure

The paper structure is as follows. Section \ref{Sec:80211ax} introduces how IEEE 802.11ax handles MU transmissions. Section \ref{Sec:SystemModel} presents the system model considered in this paper, including the main assumptions made. Section \ref{Sec:Analysis} presents the proposed IEEE 802.11ax analytical model. Section \ref{Sec:Results} presents the results and the IEEE 802.11ax performance evaluation. Finally, a summary of the most relevant contributions of the paper are listed in Section \ref{Sec:Conclusions}.

%---------------------------------------------
%---------------------------------------------
%---------------------------------------------
%---------------------------------------------

\section{IEEE 802.11ax: MU transmissions in WLANs}\label{Sec:80211ax}

\begin{table*}[t!]
  \caption{Comparison of the main IEEE 802.11ac and IEEE 802.11ax features.}  
\scriptsize
\centering
\begin{tabular}{|p{4.5cm}|p{5.5cm}|p{5.5cm}|}
\hline
    \textbf{Feature} &
      \textbf{IEEE 802.11ac} &
      \textbf{IEEE 802.11ax}
      \\
      \hline
Supported channel widths [MHz] &
      20 , 40 , 80 , 80+80 , 160  & The same
      \\
    Sub-channelization [MHz] &
      N/A &
      2.22 , 5 , 10  
      \\
    Frequency bands [GHz] &
      5  &
      2.4  and 5 
      \\
    Modulations &
      BPSK, QPSK, 16-QAM, 64-QAM, 256-QAM &
      Adds 1024-QAM
      \\
    OFDM symbol duration [$\mu$s] &
      3.6 (GI=0.4), 4  (GI=0.8) &
      13.6  (GI=0.8), 14.4  (GI=1.6), 16 (GI=3.2)
      \\
    Spatial streams (SS) &
      Up to 8 SS at the AP, up to 4 SS at user & The same
      \\
    MU transmissions &
      DL MU MIMO &
      UL and DL MU MIMO, UL and DL OFDMA
      \\
      No. of MU-MIMO users &
      4 &
      8
      \\
      Max. A-MPDU size &
      64 MPDUs &
      256 MPDUs     
      \\
    Low-density parity check (LDPC) &
      Optional &
      Mandatory
      \\
      \hline
    \end{tabular}%

  \label{tab:phy_comparison}%
\end{table*}%

In this section we introduce the operation of OFDMA and MU-MIMO transmissions as described in the IEEE~802.11ax draft, together with other important PHY and MAC changes being introduced in comparison to the IEEE 802.11ac-2013 amendment (cf. Table \ref{tab:phy_comparison}). We have considered the documents provided by the TGax\footnote{TGax: http://www.ieee802.org/11/Reports/tgax\_update.htm}, including the recently proposed draft D2.0 \cite{IEEE80211axDraft}, as the main references to build this section.

\begin{table*}[t]
\caption{The comparison of IEEE 802.11ax and IEEE 802.11ac transmission rates given in Mb/s for a single spatial stream. For IEEE 802.11ax, the GI is 3.2 $\mu$s, and for  IEEE 802.11ac the GI is 0.8 $\mu$s was assumed, which are the largest GI values in both technologies. Respectively, the OFDM symbol duration is equal to 16 and 4 $\mu$s.}
\scriptsize
\centering
\begin{tabular}{|ccc|c|ccc|cc|cc|cc|}
    \hline
     &  &  & & \multicolumn{3}{c|}{20 MHz} & \multicolumn{2}{c|}{40 MHz}  & \multicolumn{2}{c|}{80 MHz}  & \multicolumn{2}{c|}{160 MHz} \\
    \hline \hline 
    MCS & Mod. & DCM & Coding rate & Legacy (11a) & 11ac & 11ax & 11ac & 11ax & 11ac & 11ax & 11ac & 11ax \\
    \hline
    0 & BPSK & $\surd$ & 1/4 & N/A & N/A & 3.6 & N/A & 7.3 & N/A & 15.3 & N/A & 30.6 \\    
    0 & BPSK & & 1/2 & 6 & 6.5 & 7.3 & 13.5 & 14.6 & 29.3 & 30.6 & 58.5 & 61.3 \\
    1 & QPSK & $\surd$ & 1/4 & N/A & N/A & 7.3 & N/A & 14.6 & N/A & 30.6 & N/A & 61.3 \\    
    1 & QPSK & & 1/2 & 12 & 13 & 14.6 & 27.0 & 29.3 & 58.5 & 61.3 & 117 & 122.5 \\
    2 & QPSK & & 3/4 & 18 & 19.5 & 21.9 & 40.5 & 43.9 & 87.8 & 91.9 & 175.5 & 183.8 \\
    3 & 16-QAM & $\surd$ & 1/4 & N/A & N/A & 14.6 & N/A & 29.3 & N/A & 61.3 & N/A & 122.5 \\    
    3 & 16-QAM & & 1/2 & 24 & 26 & 29.3 & 54 & 58.5 & 117 & 122.5 & 234 & 245 \\
    4 & 16-QAM & $\surd$ & 3/8 & N/A & N/A & 21.9 & N/A & 43.9 & N/A & 91.9 & N/A & 183.8 \\    
    4 & 16-QAM & & 3/4 & 36 & 39 & 43.9 & 81 & 87.8 & 175.5 & 183.8 & 351 & 367.5 \\
    5 & 64-QAM & & 2/3 & 48 & 52 & 58.5 & 108 & 117 & 234 & 245 & 468 & 490 \\
    6 & 64-QAM & & 3/4 & 54 & 58.5 & 65.8 & 121.5 & 131.6 & 263.3 & 275.6 & 526.5 & 551.3 \\
    7 & 64-QAM & & 5/6 & N/A & 65 & 73.1 & 135 & 146.3 & 292.5 & 306.3 & 585 & 612.5 \\
    8 & 256-QAM & & 3/4 & N/A & 78 & 87.8 & 162 & 175.5 & 351 & 367.5 & 702 & 735 \\
    9 & 256-QAM & & 5/6 & N/A & N/A N/A & 97.5 & 180 & 195 & 390 & 408.3 & 780 & 816.6 \\
    10 & 1024-QAM & & 3/4 & N/A & N/A & 109.7 & N/A & 219.4 & N/A & 459.4 & N/A & 918.8 \\   
    11 & 1024-QAM & & 5/6 & N/A & N/A & 121.9 & N/A & 243.8 & N/A & 510.4 & N/A & 1020.8 \\   
    \hline
\end{tabular}

    \label{tab:throughput}
\end{table*}

%---------------------------------------------
%---------------------------------------------

\subsection{Physical Layer}

Similarly to IEEE 802.11ac, the IEEE 802.11ax PHY layer is based on OFDM. However, in contrast to IEEE 802.11ac in which each 20 MHz channel is divided into 64 subcarriers, IEEE 802.11ax is based on a 256-tone OFDM scheme.
The increase in the number of subcarriers is proportional to the increase in the OFDM symbol duration (from the maximum of 4 $\mu$s used in IEEE 802.11ac to the maximum of 16 $\mu$s used in IEEE 802.11ax) and guard interval (GI) duration (legacy 0.8 $\mu$s, and new 1.6 $\mu$s and 3.2 $\mu$s are supported). In terms of available transmission rates, both amendments are almost equivalent (see Table \ref{tab:throughput}). However, the use of longer OFDM symbols allows for larger coverage areas as the system becomes more robust to propagation effects, and longer GIs decrease inter-symbol interference~\cite{CISCO6th}. Note that in IEEE 802.11ax, dual carrier modulation (DCM) is used to add further protection against channel errors.

Additionally, IEEE 802.11ax keeps the same channelization as IEEE 802.11ac, i.e., 20 MHz, 40 MHz, 80 MHz, non-contiguous 80+80 MHz, and contiguous 160 MHz channels are supported. However, it extends the current OFDM scheme to multiplex several users simultaneously in the frequency domain. To this end, it introduces UL and DL OFDMA transmissions and supports additional $2.22$ MHz, $5$ MHz and $10$ MHz sub-channel widths. OFDMA sub-channels are composed of groups of subcarriers called resource units (RUs). According to the IEEE 802.11ax amendment, the maximum number of users that can be multiplexed using 2.22 MHz sub-channels in a 20 MHz channel is 9 and in a 160 MHz channel is 74. 

Furthermore, IEEE 802.11ax, similar to IEEE 802.11ac, assumes a maximum number of eight antennas at the APs and four at the user stations. However, it extends the maximum number of MU-MIMO transmissions allowed from $4$ to $8$, in both downlink and uplink directions, with up to four SU-MIMO spatial streams per user station. It is important to remark that with IEEE 802.11ax OFDMA, both DL and UL MU-MIMO transmissions can be performed along with OFDMA in RUs greater than or equal to 106 subcarriers. In this way, IEEE 802.11ax provides simultaneously user multiplexing in the spatial and frequency domains. 

The IEEE 802.11ax PHY layer includes the following additional improvements in comparison to IEEE 802.11ac: higher-order modulations (including 1024-QAM which helps to achieve 1 Gb/s rate per spatial stream) and mandatory low density parity check (LDPC) codes (with up to $1.5$-$2$ dB gain compared with traditionally used convolutional codes \cite{gong2015advanced}). Given the maximum transmission power of a node operating in the ISM band is fixed, the combined use of denser constellations and improved channel coding allows for higher transmission rates and coverage ranges.

%---------------------------------------------
%---------------------------------------------

\subsection{MAC Layer}

The enhanced distributed channel access (EDCA) mechanism is used in IEEE 802.11ax for channel contention at both APs and user stations. EDCA provides traffic differentiation by considering multiple access categories (ACs). In EDCA, each AC operates as an independent EDCA function (EDCAF) instance with different channel access parameters (CW$_{\min}$, CW$_{\max}$, AIFS). In case when multiple EDCAF instances in the same node select the same backoff counter value, EDCA defines an internal collision resolution mechanism, which guarantees the transmission from the highest priority AC involved in it.
 
To support MU transmissions, IEEE 802.11ax introduces several new control frames and procedures to the MAC layer operation: $i$) trigger-based UL MU transmissions, $ii$) MU request to send/clear to send (MU-RTS/CTS) procedure, and $iii$) multi-station block acknowledgment (MS-BACK) procedure. The details of the new procedures are the following:

%The format of the Trigger, MU-RTS and MS-BACK frames is shown in Figure \ref{Fig:Figure5}.

\begin{itemize}
    \item \textbf{Trigger-based (TB) UL MU transmission}: To request a group of user stations to perform an UL MU transmission, the AP transmits a frame called \emph{trigger}. A trigger frame contains the following information: $i$) list of user stations involved in the transmission, and $ii$) user-specific information (e.g., RU and spatial stream allocation, modulation and coding scheme). User stations, after receiving this frame, start to transmit in the assigned resources.
    \item \textbf{MU-RTS/CTS procedure}: In order to protect large MU UL and DL transmissions, the AP initiates a MU transmission sending a MU-RTS trigger frame\footnote{Note that the MU-RTS trigger frame is just a variant of the basic trigger frame.}, which includes information about user stations involved in the upcoming MU transmission, and informs about the width of the primary channels of the expected CTS frame. User stations reply transmitting simultaneously the same CTS frame on their primary channels (i.e., one or more 20~MHz channels). In case the AP is able to decode the CTS, it proceeds with the MU data transmission.  
    \item \textbf{MS-BACK procedure}: In order to reduce the large overhead required to individually acknowledge all UL MU transmissions, instead of transmitting an independent block ACK to each station, a single MS-BACK frame that aggregates all the required information for the transmitting stations is sent. 
\end{itemize}

%---------------------------------------------
%---------------------------------------------

\subsection{Data Transmissions}

IEEE 802.11ax supports both single user (SU) and MU transmissions. MU transmissions can be allocated by the AP in both UL and DL directions. The positive side of this approach is that since the AP has a complete view of the state of each user station, it can select the best stations at every transmission, which is especially relevant to achieve efficient MU transmissions. The negative side is that the AP needs to collect channel and buffer status information from user stations, which may cause significant overhead. Therefore, a trade-off exists between the amount and the rate at which such information is required, and the overhead it causes. 

\begin{figure}[t!!!!!!!!]
\centering
\epsfig{file=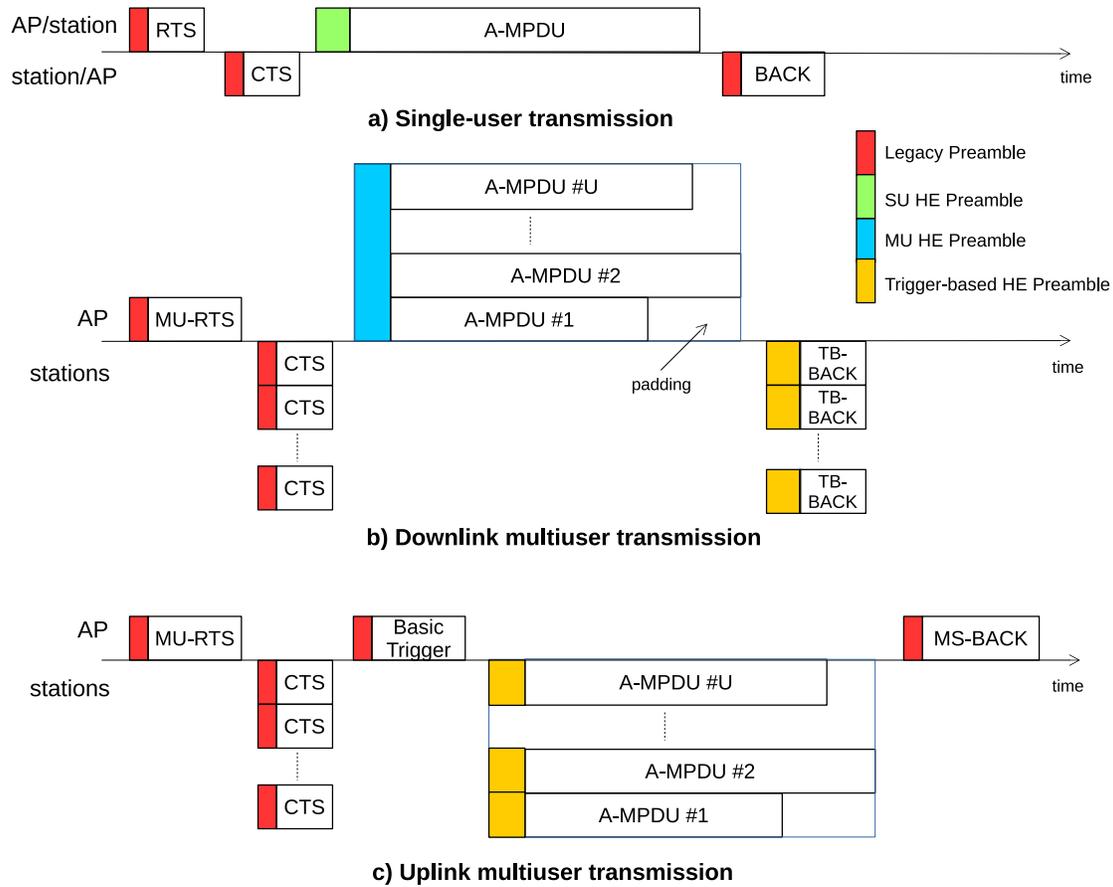,width=0.9\columnwidth,angle=-0}
\caption{SU and MU transmissions using OFDMA and MU-MIMO. Note that control frames such as the RTS, MU-RTS, Trigger, CTS, Block ACKs, MS-BACKs and NDPA transmitted in legacy mode are duplicated in every $20$~MHz sub-channel. Single and multiuser data A-MPDUs are transmitted in HE mode using the allocated RUs.} 
\label{Fig:Figure1}
\end{figure}

%---------------------------------------------

\subsubsection{SU Transmissions}

SU transmissions are performed over the entire channel width and they engage all available spatial streams in SU-MIMO mode. The standard RTS/CTS procedure can additionally be used to reserve the desired channel width, and frame aggregation can be enforced to improve transmission efficiency (cf. Figure \ref{Fig:Figure1}(a)). 

%---------------------------------------------

\subsubsection{AP-initiated MU Transmissions}

The AP may schedule (i) DL MU transmissions (Figure \ref{Fig:Figure1}(b)) and (ii) UL MU transmissions (Figure \ref{Fig:Figure1}(c)). For DL MU transmissions, since the AP is the initiator of the transmission, user selection and resource distribution between different stations do not require any further signaling mechanism. However, for UL MU transmissions, user stations are scheduled by the AP to start their simultaneous transmissions by using a trigger frame.
 
Three different types of MU transmissions are considered in IEEE 802.11ax: OFDMA, MU-MIMO, and joint MU-MIMO and OFDMA transmissions:
\begin{itemize}
    \item \textbf{MU-MIMO}: Using multiple antennas at the AP, several beams can be created in the DL direction in order to transmit data  to multiple user stations. In the UL direction, the different signals received by the antennas can also be used to separate the data sent by multiple user stations. In both cases, CSI is required at the AP, including the channel signatures of each station in both UL and DL directions. While for MU UL transmissions it can be estimated from the transmitted PHY layer preambles, to enable MU DL transmissions the stations must explicitly transmit the CSI to the AP using the channel sounding mechanism described in Section \ref{Sub:sounding}.
    \item \textbf{MU OFDMA}: The channel width is split in different sub-channels, called RUs, and allocated to different user stations. %By splitting the channel width in multiple RUs, multiple user stations can be simultaneously served by allocating each of them to a different RU.
    \item \textbf{Joint MU-MIMO and OFDMA}: In RUs larger than or equal to 106 subcarriers, MU-MIMO transmissions can also be performed, allowing to multiplex several stations at the same time over the same RU.
\end{itemize}

Spatial streams not used for MU-MIMO transmissions in a given RU, can also be used to increase the number of spatial streams allocated to individual user stations in SU-MIMO mode. 

%---------------------------------------------

\subsection{Channel and Buffer Status Information} \label{Sub:sounding}

\subsubsection{Channel State Information Acquisition}

\begin{figure}[t!!!!!!!!]
\centering
%\psfrag{Tcsbi}[][][0.8]{$T_{\text{csbi}}$}
%\psfrag{lambda}[][][0.8]{$1/\lambda_{\text{csbi}}$}
\epsfig{file=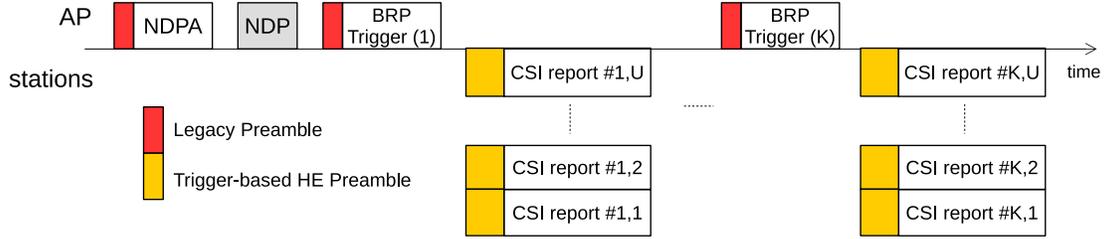,width=0.9\columnwidth,angle=-0}
\caption{CSI exchange for IEEE 802.11ax which takes advantage of the UL MU transmissions. In this figure all user stations sounded are grouped in K groups, where the user stations in each group transmit the CSI report simultaneously to the AP.} \label{Fig:CSBI11ax}
\end{figure}

The AP must know the CSI of each user station included in a MU transmission in order to create multiple beams in the DL, and to separate the multiple received streams in the UL. The explicit channel sounding mechanism presented in the IEEE 802.11ax draft is depicted in Figure \ref{Fig:CSBI11ax}. In order to achieve channel sounding, the AP sends a null data packet announcement (NDPA) frame followed by a null data packet (NDP). Additionally, in order to solicit the feedback from user stations it transmits a Beamforming Report (BRP) trigger frame.  As a response to the BRP trigger frame, stations send their CSI reports at the same time by using the new UL MU capabilities. Note that multiple trigger rounds may be necessary until all CSI reports are collected if there are more stations than the number of supported MU transmissions. 

Since the instantaneous CSI may change very fast, such a procedure must be carried periodically, which may result in a large overhead. Moreover, since there will be some delay from the time between the CSI was acquired until a station is scheduled, the CSI information collected may be inaccurate, thus also causing inefficiencies in both the user station selection and the MU beamforming, which overall may result in a low WLAN performance.

%---------------------------------------------

\subsubsection{Buffer Status Information (BSI) Acquisition}

IEEE 802.11ax introduces two complementary mechanisms for the stations to send buffer status reports (BSRs), i.e., information about the buffer status of each station, to an AP for assisting it to schedule UL MU transmissions: 

\begin{itemize}
    \item \textbf{Solicited BSR}: each station explicitly delivers its BSRs in any frame sent to the AP as a response to a BSR Poll (BSRP) trigger frame send by the AP.
    \item \textbf{Unsolicited BSR}: user stations implicitly report their BSRs in the QoS Control field and/or BSR control field of frames sent to the AP.
\end{itemize}

The use of the solicited BSR scheme comes at a cost of higher control overhead. However, it provides timely and accurate information to the AP about the current state of the stations' buffers, which should result in more efficient UL MU scheduling decisions. A possible solution to improve the efficiency of the solicited BSR mechanism is to integrate it into the periodic CSI sounding.

%---------------------------------------------
%---------------------------------------------
%---------------------------------------------
%---------------------------------------------

\section{System Model} \label{Sec:SystemModel}

\begin{figure}[ht!!!!!!!!]
\centering
\epsfig{file=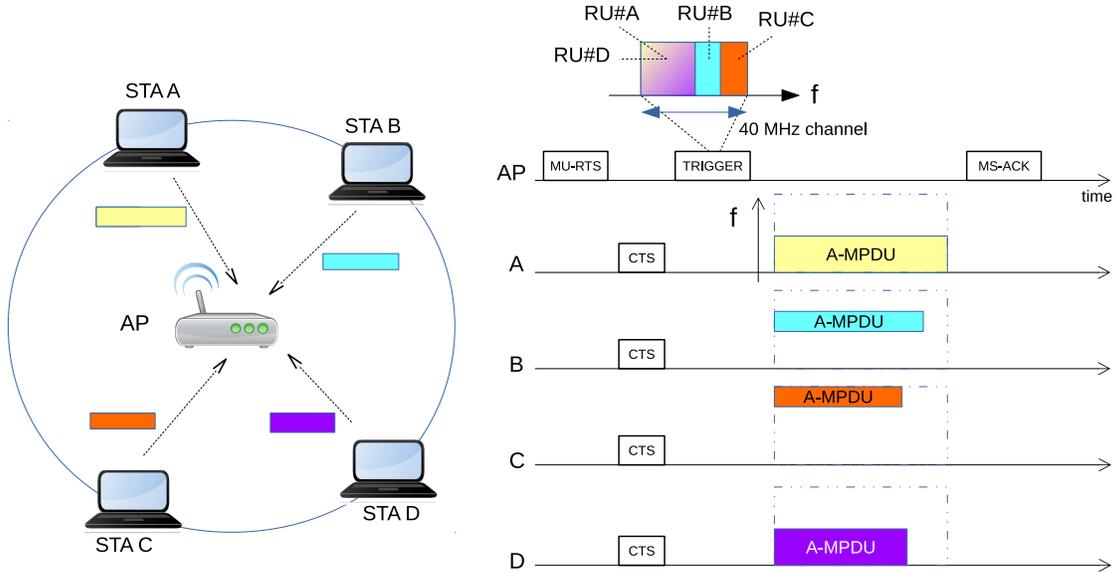,width = 0.9\columnwidth,angle=-0}
\caption{The considered scenario. It shows an example of an UL transmission where stations A and D share the same RU using MU-MIMO, while stations B and C are allocated to independent RUs}.
\label{Fig:Scenario}
\end{figure}

We consider an IEEE 802.11ax WLAN that consists of a single AP and $N$ user stations (Figure \ref{Fig:Scenario}). All stations are within the data communication range of the AP and of all other stations (i.e., there are no hidden stations), they are able to transmit and receive data using the same modulation and coding scheme, and they have exactly the same MU-MIMO and OFDMA capabilities. Additionally, we assume perfect PHY channel conditions and concentrate mostly at the MAC layer. Table \ref{Tbl:nomenclature} summarizes the notation used in this paper.

\begin{table}[t!!!!]
    \caption{Notation used in the IEEE 802.11ax analysis.} 
    \small
    \centering
    \begin{tabular}{cp{10cm}}
        \hline
        \textbf{Parameter} & \textbf{Description} \\
        \hline
        $L_{D}$ & Frame size in bits\\
        $N_a$ & Number of aggregated frames in an A-MPDU \\
	$N$ & Number of IEEE 802.11ax stations\\
%        \hline
        CW$_{\min}$ (BE) & Minimum contention window value for the AC BE\\
        CW$_{\max}$ (BE) & Maximum contention window value for the AC BE \\
        AIFS (BE) & AIFS length for AC BE \\
        AIFS$_{\text{csi}}$ & AIFS length for high priority CSI AC \\        
        SIFS & Short IFS  \\
        $\sigma$ & OFDM symbol duration \\
        $T_e$ & Empty backoff slot \\ 
 %       \hline
        $M_{\text{ap}}$ & Number of antennas at the AP  \\
        $M_{\text{sta}}$ & Number of antennas at user stations  \\
        $B$ & Channel width \\
        $N_{\text{ru}}$ & Number of RUs in a MU transmission \\
        $B_{\text{ru}}$ & Minimum channel width (minimum OFDMA sub-channel width) \\     
        $Y_m$ & Modulation used in bits/symbol \\
        $Y_c$ & Coding rate used \\
        $Y_{\text{sc}}(B_{\text{ru}})$ & Number of data sub-carriers in $B_{\text{ru}}$ \\        
        $r(V_s,B_{\text{ru}})$ & Transmission rate \\
        $V_u$ & Number of users included in a MU transmission \\
        $V_b$ & Number of OFDMA sub-channels used in a MU transmission \\ 
        $V_m$ & Number of MU-MIMO spatial streams in each RU \\
        $V_s$ & Number of SU-MIMO spatial streams per station \\
        $B_{\text{ru}}$ & Bandwidth of RU (OFDMA sub-channel) \\     
 %       \hline
        $\alpha$ & Fraction of SU DL transmissions  \\
        $\beta$ & Fraction of MU DL transmissions \\
        \hline
    \end{tabular}

    \label{Tbl:nomenclature}
\end{table}

We fix the minimum RU size to 242 data sub-carriers, which corresponds to a 20 MHz channel. In each RU, up to $M_{\text{ap}}$ MU-MIMO streams can be allocated. Therefore, with a $B=160$ MHz channel and $M_{\text{ap}}=8$ antennas, up to $64$ single stations can be multiplexed allocating a single spatial stream per station.  

A full-buffer traffic model is assumed for the AP and the stations, i.e., they have always a MAC frame ready for transmission. At each transmission, $N_{a}$ MAC frames of $L_D$-bits long are aggregated and sent following the A-MPDU packet aggregation scheme. 

EDCA is used to access the wireless channel, though only the best effort (BE) access category (AC) is considered to be active. Following the EDCA operation, when the AP backoff counter reaches zero, it starts a SU or a MU transmission with probability $\alpha$ and $1-\alpha$, respectively. Moreover, MU transmissions are DL with probability $\beta$ and UL with probability $1-\beta$. Similarly, when a user station finishes its backoff countdown, it always initiates a SU UL transmission. Unsolicited BSR is used by the stations to deliver the buffer status information to the AP.

For each MU transmission, the AP selects $V_u \leq N$ stations randomly. When $N \geq M_{\text{ap}}$, the value of $V_u$ is set as the largest value multiple of $M_{\text{ap}}$ that results in an even distribution of all the transmission resources between all selected stations (i.e., we consider that all RUs allocated have the same width ($B_{\text{ru}}$), the same number of stations is spatially multiplexed in each RU, and the same number of SU spatial streams are assigned to each station). Otherwise, the AP selects all available stations, i.e., $V_u=N$. 

Given the $V_u$ stations, the number of RUs allocated is given by $N_{\text{ru}}= \left \lceil\frac{V_u}{M_{\text{ap}}}\right \rceil$, with $V_m=\frac{V_u}{N_{\text{ru}}}$ the number of stations allocated to each RU of width $B_{\text{ru}}=\frac{B}{N_{\text{ru}}}$. The value of $V_s$, i.e., the number of spatial streams allocated to a single user, is set after assigning the MU spatial streams, and is given by $V_s=\min \left( M_{\text{sta}},\left \lfloor  \frac{M_{\text{ap}}}{V_m} \right \rfloor \right)$. For example, if $B=160$ MHz, $M_{\text{ap}}=6$, and $N=40$ stations, we will select $V_u=24$ stations, which use $N_{\text{ru}}=4$ RUs of $B_{\text{ru}}=40$ MHz each, and $V_m=6$ stations are multiplexed in each RU using MU-MIMO, with a single spatial stream ($V_s=1$) allocated to each station in SU-MIMO mode. SU transmissions, in both DL and UL, use all available bandwidth, $B$, and all available spatial streams in the SU mode, i.e., $V_s=\min \left( M_{\text{sta}},M_{\text{ap}}\right)$. 

For those frames transmitted following one of the available HE modes (i.e., SU, MU or TB), the number of transmitted bits per OFDM symbol in a RU is given by $r(V_s,B_{\text{ru}})=V_s Y_m Y_c Y_{\text{sc}}(B_{\text{ru}})$, where $Y_m$ is the number of bits per symbol of the constellation used, $Y_c$ is the coding rate, and $Y_{\text{sc}}(B_{\text{ru}})$ is the number of data subcarriers in $B_{\text{ru}}$\footnote{Note that for SU transmissions, $B_{\text{ru}}$ corresponds to the full channel width, i.e., $B_{\text{ru}}=B$.}. Differently, control frames are transmitted in legacy mode using the basic rate of $r_{\text{legacy}}=24$ bits per OFDM symbol (i.e., 6 Mb/s, MCS $0$), with all control frames duplicated in every 20 MHz sub-channel when wider channels are used, using a single spatial stream.

The AP periodically requests the CSI from the stations at a rate $\lambda_{\text{csi}}$ requests/second. The CSI procedure has a duration of $T_{\text{csi}}$ seconds. 
To achieve a higher priority than other transmissions for the channel sounding, we consider that the AP uses a high priority AC for control and management frames that is configured with the minimum feasible AIFS value and CW$_{\min}=0$. These values guarantee that when the AP decides to start the channel sounding process it will get access to the channel as soon as the channel is detected idle, and without colliding with any UL transmission\footnote{In dense scenarios, this approach may cause inter-WLAN collisions when two APs decide to initiate the CSI operation exactly at the same time. Therefore, appropriate overlapping basic service set (OBSS) management mechanisms are needed, though, this is out of the scope of this paper.}. Therefore, the available time for data transmissions is limited to $T_{\text{data}}=\frac{1}{\lambda_{\text{csi}}}-T_{\text{csi}}$. 

Finally, in order to provide a clear picture of the link-layer IEEE 802.11ax saturation throughput for a single WLAN, we consider an ideal channel without errors. Moreover, there is no capture effect, and in all cases, collisions result in the loss of all transmitted frames. 

%---------------------------------------------

\section{Saturation Throughput Model}\label{Sec:Analysis}

In this section we introduce the analytical model used to calculate the expected saturation throughput under assumptions listed in Section \ref{Sec:SystemModel}. Our analysis is based on the well-known Bianchi's IEEE 802.11 DCF model~\cite{bianchi2000performance}, which has been proven accurate for the analysis of WLAN performance. However, in comparison to Bianchi's model, the following parts have been extended to capture the IEEE 802.11ax characteristics: different types and lengths of transmissions (cf. Section \ref{sec:td}), additional overhead introduced by the channel sounding procedure (cf. Section \ref{sec:tdcs}), different probabilities of successful transmissions and collisions in the presence of different types of SU and MU transmissions (cf. Section \ref{sec:prob}), and separate performance metrics for UL and DL transmissions (cf. Section \ref{sec:perf}).

%--------------------------------------------

\subsection{SU and MU Transmission Duration} \label{sec:td}

%--------------------------------------------

\subsubsection{Successful transmissions}

Respectively, the duration times of SU ($T_{\text{su}}$) and MU (downlink $T_{\text{mu,d}}$ and uplink $T_{\text{mu,u}}$) transmissions, understood as the time in which the channel is detected as busy by the non-transmitting stations, are given by
\begin{align} \label{Eq:TSUandMU}
T_{\text{su}}(V_s,B) & = T_{\text{RTS}} + T_{\text{SIFS}} + T_{\text{CTS}} + T_{\text{SIFS}} + T^{\text{D}}_{\text{su}}(V_s,B) + T_{\text{SIFS}} + T_{\text{BACK}} + \text{AIFS}, \\ 
T_{\text{mu,d}}(V_u,V_s,B_{\text{ru}}) & =T_{\text{MU-RTS}}(V_u) + T_{\text{SIFS}} + T_{\text{CTS}}  + T_{\text{SIFS}} + T^{\text{D}}_{\text{mu}}(V_s,B_{\text{ru}}) +  T_{\text{SIFS}}  + T_{\text{BACK}}+\text{AIFS}, \text{ and}   \nonumber \\
T_{\text{mu,u}}(V_u,V_s,B_{\text{ru}}) & =  T_{\text{MU-RTS}}(V_u) + T_{\text{SIFS}} + T_{\text{CTS}} + T_{\text{SIFS}} + T_{\text{trigger}}(V_u) + T_{\text{SIFS}} +  \nonumber \\+ & T^{\text{D}}_{\text{mu,u}}(V_s,B_{\text{ru}})+ T_{\text{SIFS}}  + T_{\text{MS-BACK}}(V_u) + \text{AIFS},    \nonumber
\end{align}
where the duration of the data frame is the following\footnote{In case $Na=1$, the term $N_a (L_{\text{MD}}+L_{\text{MH}}+L_{\text{D}})$ is reduced to $(L_{\text{MH}}+L_{\text{D}})$.} 
\begin{small}
\begin{align}\label{Eq:TxDurations}
T^D_{\text{su}}(V_s,B) & = T_{\text{PHY-HE-SU}} +  \left\lceil \frac{L_{\text{SF}} + N_a (L_{\text{MD}}+L_{\text{MH}}+L_{\text{D}}) + L_{\text{TB}}}{r(V_s,B)}\right\rceil \sigma, \nonumber \\ 
T^D_{\text{mu,d}}(V_s,B_{\text{ru}}) & = T_{\text{PHY-HE-MU}} +  \left\lceil \frac{L_{\text{SF}} + N_a (L_{\text{MD}}+L_{\text{MH}}+L_{\text{D}}) + L_{\text{TB}}}{r(V_s,B_{\text{ru}})}\right\rceil \sigma, \nonumber \\ 
T^D_{\text{mu,u}}(V_s,B_{\text{ru}}) & = T_{\text{PHY-HE-TB}}+ \left\lceil \frac{L_{\text{SF}} + N_a (L_{\text{MD}}+L_{\text{MH}}+L_{\text{D}}) + L_{\text{TB}}}{r(V_s,B_{\text{ru}})}\right\rceil \sigma, \nonumber
\end{align}
\end{small}
and the duration of control frames is
\begin{small}
\begin{align}
T_{\text{RTS}} & = T_{\text{PHY-legacy}} + \left\lceil \frac{L_{\text{SF}} + L_{\text{RTS}} + L_{\text{TB}}}{r_{\text{legacy}}}\right\rceil \sigma_{\text{legacy}} \nonumber \\
T_{\text{MU-RTS}}(V_u) & = T_{\text{PHY-legacy}} + \left\lceil \frac{L_{\text{SF}} + L^{\text{MU-RTS}}_{\text{trigger}}(V_u) + L_{\text{TB}}}{r_{\text{legacy}}}\right\rceil \sigma_{\text{legacy}} \nonumber \\
T_{\text{CTS}} & = T_{\text{PHY-legacy}} + \left\lceil \frac{L_{\text{SF}} + L_{\text{CTS}} + L_{\text{TB}}}{r_{\text{legacy}}}\right\rceil \sigma_{\text{legacy}}  \nonumber \\ 
T_{\text{trigger}}(V_u) & = T_{\text{PHY-legacy}} + \left\lceil \frac{L_{\text{SF}} + L^{\text{BASIC}}_{\text{trigger}}(V_u) + L_{\text{TB}}}{r_{\text{legacy}}}\right\rceil \sigma_{\text{legacy}}  \nonumber \\
T_{\text{BACK}} & = T_{\text{PHY-legacy}} + \left\lceil \frac{L_{\text{SF}} + L_{\text{BACK}} + L_{\text{TB}}}{r_{\text{legacy}}}\right\rceil \sigma_{\text{legacy}}  \nonumber \\
T_{\text{TB-BACK}}(V_s,B_{\text{ru}}) & = T_{\text{PHY-HE-TB}} + \left\lceil \frac{L_{\text{SF}} + L_{\text{BACK}} + L_{\text{TB}}}{r(V_s,B_{\text{ru}})}\right\rceil \sigma  \nonumber \\
T_{\text{MS-BACK}}(V_u) & = T_{\text{PHY-legacy}} + \left\lceil \frac{L_{\text{SF}} + L_{\text{MS-BACK}}(V_u) + L_{\text{TB}}}{r_{\text{legacy}}}\right\rceil \sigma_{\text{legacy}} \nonumber 
\end{align}
\end{small}
where $\sigma$ is the duration of an OFDM symbol (cf. Table \ref{tab:phy_comparison}) and $N_a$ is the number of frames aggregated in an A-MPDU. The explanations and the lengths of the used variables are given in Table \ref{Tbl:11axParameters}.

\begin{table}
	\caption{IEEE 802.11ax frame lengths, PHY preamble durations, and other parameters}
	\centering
	\begin{tabular}{lcc}
	Explanation & Name of the variable & Value \\
	\hline \hline
	Legacy preamble & $T_{\text{PHY-legacy}}$ & 20 $\mu$s \\
	HE Single-user preamble & $T_{\text{PHY-HE-SU}}$  & 164 $\mu$s \\
	HE Multiuser preamble & $T_{\text{PHY-HE-TB}}$ & 228 $\mu$s \\
	HE Trigger-based preamble & $T_{\text{PHY-HE-MU}}$ & 168 $\mu$s \\
	\hline	
	Service field & $L_{\text{SF}}$ & 16 bits \\
	MPDU Delimiter & $L_{\text{MD}}$ & 32 bits \\
	MAC header & $L_{\text{MH}}$ & 320 bits \\%\textcolor{red}{why? In Figure 9-1: MAC frame format 802.11-2016 it is 32 header + 4 fcs  = 36 Bytes = 288 bits } \\
	Tail bits & $L_{\text{TB}}$ & 18 bits \\ %\textcolor{red}{why not 6?} \\
	\hline	
	Request To Send (RTS) &  $L_{\text{RTS}}$ & 160 bits \\
	Clear To Send (CTS) &  $L_{\text{CTS}}$ & 112 bits \\
	Multi-user RTS trigger & $L^{\text{MU-RTS}}_{\text{trigger}}(V_u)$ & $224 + 40 \cdot V_u$ bits (largest case) \\
	Basic trigger & $L^{\text{BASIC}}_{\text{trigger}}(V_u)$ & $224 + 48 \cdot V_u$ bits (largest case) \\
	Beamforming Report Poll (BRP) trigger & $L^{\text{BRP}}_{\text{trigger}}(V_u)$ & $224 + 48 \cdot V_u$ bits (largest case) \\
	Block ACK (BACK) & $L_{\text{BACK}}$ & 256 bits \\
	Multi-station BACK (MS-BACK) & $T_{\text{MS-BACK}}(V_u)$ & $176+288 \cdot V_u$ \\	
	Null Data Packet Announcement (NDPA) & $L_{\text{NDPA}}(N)$ & $168 + 32 N$ bits \\
	Null Data Packet (NDP) & $T_{\text{NDP}}$ & 168 $\mu$s\\
	Beamforming Report (BREPORT) & $L_{\text{BREPORT}}(B)$ & $64+N_{\text{ang}} \frac{(b_{\psi}+b_{\phi})}{2} \frac{Y_{\text{sc}}(B)}{N_{\text{sg}}}$\\
 	\hline
	\end{tabular}

	\label{Tbl:11axParameters}
\end{table}

\subsubsection{Collisions}

The duration of a collision is given by the largest transmission involved in it. We must consider the following two cases:

\begin{enumerate}
    \item A SU transmission from the AP, or from one of the stations, collides with one or more SU transmissions from stations. In this case, the collision event has a duration of $T_{\text{c,su}}=T_{\text{RTS}}+T_{\text{SIFS}}+T_{\text{ack-time-out}}$, where the ACK time-out is the time which a station waits to restart its normal activity. 
    \item A DL or UL MU transmission collides with one or more SU transmissions from the stations. In this case, the duration of the collision is given by  $T_{\text{c,mu}}(V_u)=T_{\text{MU-RTS}}(V_u)+T_{\text{SIFS}}+T_{\text{ACK-time-out}}$, as the MU-RTS duration is larger than the legacy RTS sent by the station in case of the SU transmission. Note, in addition, that in this case the collision duration depends on the number of stations included in the MU transmission.
\end{enumerate}
The value of the $T_{\text{ACK-time-out}}$ is set to $T_{\text{ack-time-out}}=T_{\text{CTS}}+T_{\text{AIFS}}$ to guarantee all nodes re-start their backoff counter at the same time. 

%---------------------------------------------
%---------------------------------------------

\subsection{Duration of the Channel Sounding Procedure}
\label{sec:tdcs}

The duration of the CSI sounding procedure, $T_{\text{csi}}$, in IEEE 802.11ax follows the description given in Figure~\ref{Fig:CSBI11ax}, and it is given by: 
\begin{align} 
T_{\text{csi}}(N,B) & =  T_{\text{NDPA}}(N) + T_{\text{SIFS}} + T_{\text{NDP}} +   K \left(T_{\text{SIFS}}  + T^{\text{BRPOLL}}_{\text{trigger}}(N) + T_{\text{SIFS}} + T_{\text{BREPORT}}(B) \right ) 
+ \text{AIFS}_{\text{CSI}},
\end{align}
where $\text{AIFS}_{\text{CSI}}$ is the AIFS duration of the highest priority AC used for the channel sounding. $K$ is the number of groups in which stations are allocated to transmit the CSI simultaneously, as shown in Figure~\ref{Fig:CSBI11ax}. $ T^{\text{BRPOLL}}_{\text{trigger}}$ and $T_{\text{BREPORT}}$ stand for the Beamforming Report Poll Trigger and the Beamforming Report, respectively. 

% =\left \lceil \frac{N}{B/20} \right \rceil

The duration of the Null Data Packet Announcement (NDPA), Null Data Packet (NDP) and CSI reports are the following:
\begin{align}
    T_{\text{NDPA}}(N) & = T_{\text{PHY-legacy}} + \left \lceil \frac{L_{\text{SF}}+L_{\text{NDPA}}(N) + L_{\text{TB}}}{r_{\text{legacy}}} \right \rceil \sigma_{\text{legacy}}, \nonumber\\
    T_{\text{NDP}} & = 168~\mu\text{s},~\text{and} \nonumber\\
    T^{\text{BRPOLL}}_{\text{trigger}}(N) & = T_{\text{PHY-legacy}} + \left \lceil \frac{L^{\text{BRPOLL}}_{\text{trigger}}(N)}{r_{\text{legacy}}} \right \rceil \sigma_{\text{legacy}}, \nonumber\\
    T_{\text{BREPORT}}(B) &= T_{\text{PHY-HE-TB}} + \left \lceil\frac{L_{\text{SF}}+L_{\text{MH}}+ L_{\text{BREPORT}}(B)+ L_{\text{TB}}}{r(1,B_{\text{ru}})}\right \rceil \sigma, \nonumber
\end{align}
where the size of the CSI reports depends on the number of angles used to estimate the channel matrix for each subcarrier ($N_{\text{ang}}$), the number of bits used to quantize these angles ($b_{\psi}+b_{\phi}$), the total number of subcarriers ($Y_{sc}(B)$), and the number of subcarriers grouped ($N_{\text{sg}}$). Further details can be found in Table~\ref{Tbl:11axParameters}.

%---------------------------------------------
%---------------------------------------------

\subsection{Probability of Successful Transmissions and Collisions}
\label{sec:prob}

The temporal evolution of a WLAN can be considered slotted under saturation conditions if all nodes are able to decrease their backoff counter at exactly the same time, with the duration of a backoff slot defined as the time between two consecutive backoff decrements by a station. This time may vary depending on the number of transmissions initiated at each slot. Therefore, the duration of each slot depends on the probability that nodes transmit in it or remain idle, and if it contains a successful transmission or a collision. When the number of transmissions in a slot is zero, we refer to it as an empty slot with the duration of $T_e$.

Under these conditions, since we have only two types of nodes (AP and user stations), and all user stations operate exactly in the same way, we know from \cite{bellalta2005simple} that either the AP or a single user station transmits in a randomly chosen slot with probability 
\begin{align}\label{Eq:taus}
    \tau_{\text{ap}} &= \frac{1}{E[\Phi(\text{CW}_{\text{min,ap}},m_{\text{ap}},p_{\text{c,ap}})]+1},~\text{and}~ \nonumber \\ \tau_{\text{sta}} &= \frac{1}{E[\Phi(\text{CW}_{\text{min,sta}},m_{\text{sta}}	,p_{\text{c,sta}})]+1},
\end{align}
respectively, with $E[\Phi(\text{CW}_{\text{min}},m,p_c)]$ the expected number of backoff slots,
%\begin{align}
%    E[\Phi(\text{CW}_{\text{min}},m,p_c)]=\frac{1-p_{c}-p_{c}(2p_{c})^{m}}{1-2p_{c}}\frac{\text{CW}_{\min}}{2}-\frac{1}{2}, \nonumber    
%\end{align}
%\begin{align}
%    E[\Phi(\text{CW}_{\text{min}},m,p_c)]=\frac{1-p_{c}-p_{c}(2p_{c})^{m}}{1-2p_{c}}\frac{\text{CW}_{\min}}{2}, \nonumber    
%\end{align}
\begin{align}
    E[\Phi(\text{CW}_{\text{min}},m,p_c)]&= \sum_{i=0}^{m}{2^i\frac{\text{CW}_{\min}}{2}p_c^i(1-p_c)}+\sum_{i=m+1}^{\infty}{2^m\frac{\text{CW}_{\min}}{2}} p_c^i(1-p_c) \nonumber \\
    & = \frac{\text{CW}_{\min}}{2}(1-p_c) \sum_{i=0}^{m}{2^ip_c^i} + \frac{\text{CW}_{\min}}{2}(1-p_c) 2^m \sum_{i=m+1}^{\infty}{ p_c^i} \nonumber \\
    & = \frac{\text{CW}_{\min}}{2} \left( \frac{1-p_c - (2p_c)^{m+1} + p_c(2p_c)^{m+1}}{1-2p_c} +  p_c(2p_c)^{m} \right) \nonumber \\
    &= \frac{1-p_{c}-p_{c}(2p_{c})^{m}}{1-2p_{c}}\frac{\text{CW}_{\min}}{2}, \nonumber    
\end{align}
where $\text{CW}_{\text{min}}$ is the minimum contention window, $m$ is the number of backoff stages, and $p_c$ is the collision probability.

The probability that a transmission from the AP results in a collision is given by the probability that at least one station has also initiated a transmission at the same slot that the AP,
\begin{align}
    p_{c,\text{ap}}& =1-(1-\tau_{\text{sta}})^{N}.\nonumber
\end{align}
Similarly, the probability that a transmission from a station results in a collision is given by the probability that either another station or the AP have initiated a transmission at the same backoff slot,
\begin{align}
    p_{c,\text{sta}} &=1-(1-\tau_{\text{ap}})(1-\tau_{\text{sta}})^{N-1}. \nonumber
\end{align}

A backoff slot contains a successful transmission only if either the AP or a single station transmits in it. Since we have four types of transmissions (i.e., DL and UL SU transmissions, and DL and UL MU transmissions) we calculate the successful transmission probabilities for each type as:

\begin{enumerate}
	\item The probability that the AP initiates a DL SU transmission and none of the user stations transmit:  $a_1 = \alpha \tau_{\text{ap}}(1-\tau_{\text{sta}})^{N_{}}$.
	\item The probability that exactly a single user station transmits (i.e., it initiates an UL SU transmission) and neither the AP nor  other user stations transmit: $a_2 = N \tau_{\text{sta}}(1-\tau_{\text{ap}})(1-\tau_{\text{sta}})^{N_{}-1}$.
	\item The probability that the AP starts a DL MU transmission, and none of the user stations transmit: $a_3 = (1-\alpha) \beta \tau_{\text{ap}}(1-\tau_{\text{sta}})^{N_{}}$.
 	\item The probability that the AP starts a UL MU transmission, and none of the user stations transmit: $a_4 = (1-\alpha) (1-\beta) \tau_{\text{ap}}(1-\tau_{\text{sta}})^{N_{}}$.
\end{enumerate}

In all previous four cases there are always two conditions to meet to have a successful transmission: 1) the probability that the AP or one user station performs a transmission in a given slot, and 2) the probability that the remaining nodes do not transmit in this slot. For example, $a_3$ requires that the AP transmits ($\tau_{\text{ap}}$), the transmission is of MU type ($1-\alpha$), it is performed in the DL direction ($\beta$), and all the user stations are silent in that slot ($(1-\tau_{\text{sta}})^{N_{}}$).

The probability a slot remains 'empty' is $b_1=(1-\tau_{\text{ap}})(1-\tau_{\text{sta}})^{N_{}}$, representing the case in which neither the AP nor the stations transmit in that slot.

When more than one node starts a transmission in the same slot, there is a collision. Similarly to the case with successful transmissions, there are four situations to consider: 
\begin{enumerate}
	\item A collision between the AP when it starts a SU transmission and one or more transmissions initiated by the stations, $c_1=\alpha\tau_{ap}\left(1-(1-\tau_{sta})^{N_{}} \right)$.
	\item A collision between the DL MU transmission initiated by the AP and one or more transmissions initiated by the stations, $c_2=(1-\alpha) \beta \tau_{ap}\left(1-(1-\tau_{sta})^{N_{}} \right)$.
	\item A collision between the UL MU transmission initiated by the AP and one or more transmissions initiated by the stations,  $c_3=(1-\alpha) (1-\beta) \tau_{ap}\left(1-(1-\tau_{sta})^{N_{}} \right)$.
	\item A collision between two or more transmissions initiated by the stations, $c_4=1-a_1-a_2-a_3-a_4-b_1-c_1-c_2-c_3$, which is equivalent to sum of the probabilities of all possible combinations in which two or more user stations transmit in the same slot.
\end{enumerate}

Note that in the first three cases the probability that a backoff slot contains a collision is calculated as the probability that the AP performs a transmission multiplied by the probability that at least one station also transmits.

%---------------------------------------------

In our analysis, a fixed-point approach is used to solve the non-linear system of equations given by 
$\tau_{\text{ap}}$, $\tau_{\text{sta}}$, $p_{c,\text{ap}}$ and $p_{c,\text{sta}}$.

\subsection{UL and DL Throughput}
\label{sec:perf}

The throughput in the DL and UL directions is given by

\begin{align}
    S_d = \frac{T_{\text{data}}}{T_{\text{csi}}+T_{\text{data}}} \left( \frac{a_1 N_a L_D + a_3 V_u N_a L_D}{b_1 T_{e}+\sum_{i=1}^{4}{a_i(T_{a_i}+T_{e})}+\sum_{i=1}^{4}{c_i(T_{c_i}+T_{e})}}\right) \nonumber
\end{align}
and
\begin{align}
    S_u = \frac{T_{\text{data}}}{T_{\text{csi}}+T_{\text{data}}} \left( \frac{a_2 N_a L_D + a_4 V_u N_a L_D}{b_1 T_{e}+\sum_{i=1}^{4}{a_i(T_{a_i}+T_e)}+\sum_{i=1}^{4}{c_i(T_{c_i}+T_e)}}\right)  \nonumber
\end{align}
respectively, where  $T_{a_1}=T_{\text{su}}(V_s,B)$ is the duration of a DL SU transmission, $T_{a_2}=T_{\text{su}}(V_s,B)$ is the time of an UL SU transmission, $T_{a_3}=T_{\text{mu,d}}(V_u,V_s,B_{\text{ru}})$ is the time of a DL MU transmission, $T_{a_4}=T_{\text{mu,u}}(V_u,V_s,B_{\text{ru}})$ is the duration of an UL MU transmission,  $T_{c_1}=T_{c_4}=T_{\text{c,su}}$ are the collision duration times for SU transmissions, and $T_{c_2}=T_{c_3}=T_{\text{c,mu}}(V_u)$ are the collision duration times for MU transmissions. Notice that successful transmission and collision slots also include an empty slot, as otherwise, nodes would not be able to decrease their backoff counter. The term $\frac{T_{\text{data}}}{T_{\text{csi}}+T_{\text{data}}}$ takes into account the fraction of time used for the channel sounding. 

Once the saturation throughput is calculated, the expected service time per transmission (i.e., the time required to successfully transmit a frame) is simply given by $E[D_u]=\frac{L_D}{S_u}$ and $E[D_d]=\frac{L_D}{S_d}$, respectively for the UL and DL directions. Note that the service time includes the transmission delay, the backoff time duration and other temporal overheads.

%---------------------------------------------
 
\section{Results} \label{Sec:Results}

\begin{table}[t]
    \caption{Parameters used to obtain the numerical results.} 
    \small
    \centering
\begin{tabular}{p{8cm}p{6cm}}
        \textbf{Parameter} & \textbf{Value}\\
        \hline\hline
        $L_D$ & 12000 bits \\
        Max. A-MPDU size & 256 frames \\
        Maximum PPDU duration & 5.4884 ms \\
%        \hline
        CW$_{\min,\text{ap}}$ (BE)  & 15 \\
		CW$_{\min,\text{sta}}$ (BE)  & 15 \\
        CW$_{\max,\text{ap}}$ (BE)  & 1023 \\
		CW$_{\max,\text{sta}}$ (BE)  & 1023 \\		
        Number of backoff stages, $m=\log_2\frac{\text{CW}_{\max}}{\text{CW}_{\min}}$ (BE)  & 6 \\ %\textcolor{red}{For the case with RTS/CTS this should be set to 7}\\
        AIFS (BE) & 34 $\mu$s \\
        AIFS$_{\text{csi}}$ & 25 $\mu$s \\        
        SIFS & 16 $\mu$s \\
        $T_e$ & 9 $\mu$s \\ 
 %       \hline
 %       $\sigma$ & 16 $\mu$s \\
        $M_{\text{ap}}$ &  8  \\
        $M_{\text{sta}}$ &  4   \\
        $B$ & 160 MHz \\
		%GI, $\sigma$ & 3.2 $\mu$s, 16 $\mu$s \\
		$\sigma$ & 16 $\mu$s \\
		$\sigma_{\text{legacy}}$ & 4 $\mu$s \\
        %$B_{\min}$ & 5 MHz (i.e., 52 data subcarriers) \\        
%        $\alpha$ & 0.2 \textcolor{red}{can we present the results for variable alpha and beta?}\\
 %       $\beta$ & 0.8 \\
%         MCS index (data and control frames) & 6 ($Y_{m}=6$, $Y_{c}=3/4$) \\
         MCS index (HE frames) & 6 ($Y_{m}=6$, $Y_{c}=3/4$) \\
%		 MCS index (Legacy frames) & 0 ($Y_{m}=1$, $Y_{c}=1/2$ -- $6$ Mbps) \\
%		 $r_{\text{legacy}}$ & $1/2$ (BPSK, 1/2) -- 6 Mbps \\
%         MCS index (control information) & 6 ($Y^{\text{basic}}_m=6$, $Y^{\text{basic}}_c=3/4$ ) \\        
%        $Y^\text{basic}_m$  & 6 \
%        $Y^\text{basic}_c$ & 3/4 \\
    %    \hline
        $\lambda_{\text{csi}}$ & 20 attempts/second\\     
%        $T_{\text{csbi}}$ & $\approx 1.3+0.13\lceil\frac{N_{\text{stas}}}{M_{\text{ap}}}\rceil$ ms  \textcolor{red}{why} \boris{To avoid explaining how that is calculated. It is a lot of information to get just that. For sure, we need to explain why we use that expression}\\
		$K$ & 1 \\ % $\left \lceil \frac{N}{8} \right \rceil$ \\
        CSI sounding: $N_{\text{ang}}$, $b_{\psi}$, $b_{\phi}$, $N_{\text{sg}}$  & 56 angles, 8 bits, 8 bits, 16 subcarriers \\
        \hline\hline
    \end{tabular}

    \label{Tbl:parameters}
\end{table}

In this section we investigate the saturation throughput of an IEEE 802.11ax WLAN by applying the analysis developed in Section \ref{Sec:Analysis}. We focus our attention on understanding the throughput gains provided by the new MU transmission capabilities included in the IEEE 802.11ax draft. We also investigate the tradeoff between the efficiency and the introduced overheads when CSI is collected from many stations in order to perform DL MU transmissions. Finally, we show the benefits of reducing the stations' transmission attempts when the AP is allowed to schedule UL MU transmissions with respect to their own throughput.

To validate the correctness of the analytical model, we include the throughput values obtained by simulation. The simulator has been developed in Matlab\footnote{Since the required modules to support MU transmissions in IEEE 802.11ax are not yet available in NS3, we chose to develop a simulator from scratch to validate the correctness of the presented analysis.}, and it accurately reproduces the features described in Section \ref{Sec:80211ax}. Each simulation point has been obtained by averaging $20$ executions of $10$ seconds. The error bars show the standard deviation of the throughput values obtained at each execution. In the plots, we label with (S) the curves obtained by simulation, and with (A) the curves obtained using the analysis.

Table \ref{Tbl:parameters} shows the values of the considered parameters  to obtain the results presented in this section, unless otherwise stated. We have assumed that the CSI is requested every $50$ ms, although TGax results show that even values higher than $500$ ms may be acceptable between two consecutive CSI requests in low mobility scenarios \cite{Breit}.

%---------------------------------------------

\subsection{MU Transmissions and Frame Aggregation to Overcome WLAN Inefficiency}

\begin{figure}[t]
\centering
\subfigure[Transmission rates and SU DL throughput] {\epsfig{file=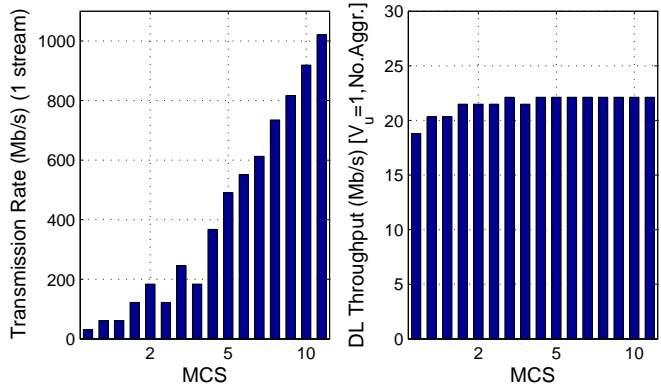,width=0.6\columnwidth,angle=-0}\label{Fig:F1_SUModes}}
%
%
%\hspace{2mm}
%
\subfigure[MU DL throughput]{\epsfig{file=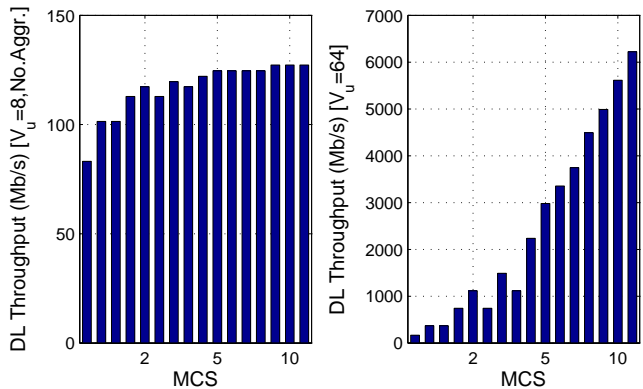,width=0.6\columnwidth,angle=-0}\label{Fig:F1_MUModes1}}
\caption{Transmission rates versus SU and MU throughput of DL transmissions when only the AP is transmitting (i.e., the AP is the only node with frames to transmit in the WLAN) calculated for each MCS (cf. Table \ref{tab:throughput}).}
\label{Fig:RatesSUandMUThroughput}
\end{figure}

Figure \ref{Fig:RatesSUandMUThroughput} shows the transmission rate and throughput for each IEEE 802.11ax MCS in different configurations. In this subsection, we consider that only the AP is transmitting, and therefore there are no collisions. The channel sounding mechanism is also disabled. The results illustrate the inefficiency of IEEE 802.11ax (and of IEEE 802.11 WLANs in general) when performing SU transmissions, and how such inefficiency can be mitigated by using frame aggregation and MU-transmissions\footnote{Since the maximum PPDU duration is limited to 5.484 ms, sharing the channel resources between different users may reduce the number of frames that can be aggregated in each A-MPDU, thus resulting in lower than expected throughput gains.}. Comparing the left-side (transmission rate) with the right-side (throughput) of Figure \ref{Fig:F1_SUModes} it is observed that transmission rates of up to 1 Gb/s result in an effective throughput of less than $25$ Mb/s. These low throughput values are the result of the high overheads included in each transmission: the backoff time, the RTS/CTS exchange done at the basic transmission rate, inter-frame spaces (AIFS and SIFS), the PHY and MAC headers, and the acknowledgment procedure, among others. For example, to transmit a single frame of $12000$ bits, it is required a total time of $542.5~\mu$s using MCS 9 ($T_s(V_s,B)$ from (\ref{Eq:TSUandMU})), which results in the throughput of $22.12$ Mb/s shown in the right side of Figure \ref{Fig:F1_SUModes}. The use of MU transmissions and frame aggregation (cf. Figure \ref{Fig:F1_MUModes1}) allows to mitigate such high inefficiency, achieving aggregate throughput values of up to $6$ Gb/s.

%---------------------------------------------
%---------------------------------------------

\subsection{Negative Effect of Collisions and CSI Requests} 

%\kks{I do not uderstand this drop in DL, if the AP is the only transmitter there are no collisions etc. This should be explained in the text.}

\begin{figure}[t]
\centering
\subfigure[DL and UL IEEE 802.11ax and IEEE 802.11ac aggregate throughput as a function of the active number of stations ($N$) in the WLAN. \break For each MU transmission, $V_u \leq N$ stations are selected following the scheduling approach described in Section \ref{Sec:SystemModel}.]{\epsfig{file=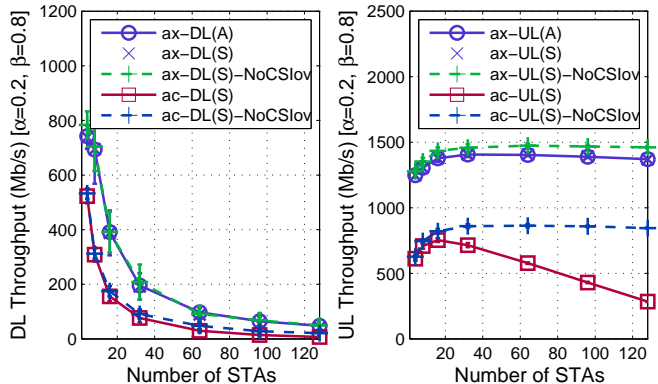,width=0.6\columnwidth,angle=-0}\label{Fig:FigureContAlfa02Beta08}}\\
%
%\hspace{2mm}
%
\subfigure[Impact of CSI overheads on throughput for a variable number of stations (IEEE 802.11ax only).  ]{\epsfig{file=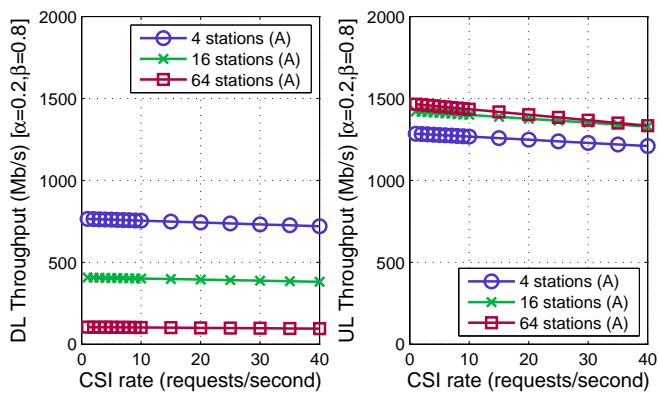,width=0.6\columnwidth,angle=-0}\label{Fig:FigureCSBIrate}}
\caption{DL and UL throughput.}
\label{Fig:Throughput11ax}
\end{figure}

\begin{figure}[t]
\centering
\subfigure[$\alpha$ parameter.]{\epsfig{file=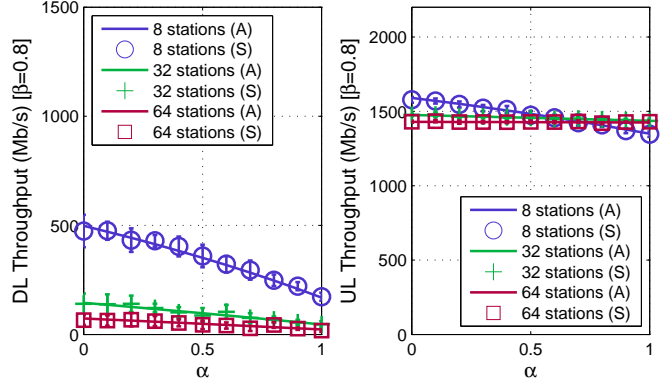,width=0.6\columnwidth,angle=-0}\label{Fig:FigureAlpha}}\\
\subfigure[$\beta$ parameter.]{\epsfig{file=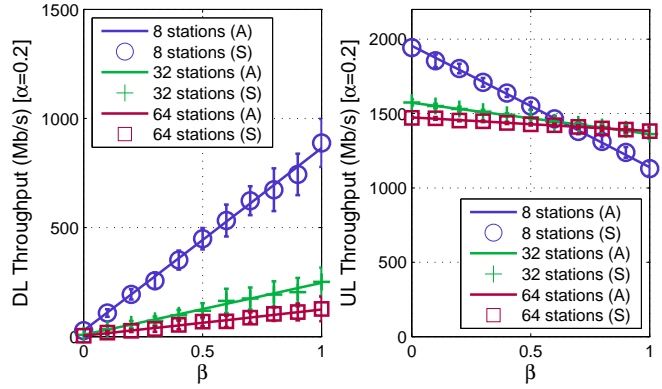,width=0.6\columnwidth,angle=-0}\label{Fig:FigureBeta}}
\caption{Effect of different $\alpha$ and $\beta$ values on the UL and DL throughput for different number of stations.}
\label{Fig:FiguAlphaBeta}
\end{figure}

In IEEE 802.11ax WLANs the presence of many active user stations allows to schedule large MU transmissions which, as we have seen before, may result in significant high throughput values. However, the presence of a large number of active stations contending for the channel may also result in collisions, thus negatively affecting the overall WLAN performance. In addition, a high number of active user stations also results in high temporal overheads due to the need to collect the CSI observed by all of them.%channel sounding procedure. 

Figure \ref{Fig:FigureContAlfa02Beta08} shows the DL and UL throughput achieved by a single WLAN when it implements either IEEE 802.11ax ($\alpha=0.2$ and $\beta=0.8$) or IEEE 802.11ac ($\alpha=0.2$ and $\beta=1$, as no UL MU transmissions are allowed) amendments. Using the same MCS, the new capabilities introduced by the IEEE 802.11ax amendment, such as increasing the number of frames that can be aggregated in a single A-MPDU (from $64$ to $256$ frames), and the support for large UL and DL MU transmissions involving many stations (from $4$, and only in the DL, to more than $64$ user stations in both directions), result in a higher throughput for both DL and UL. 

Figure \ref{Fig:FigureContAlfa02Beta08} also shows the performance gains of using the HE channel sounding mechanism. It allows user stations to send to the AP the CSI information in parallel instead of sequentially, as it is done in IEEE 802.11ac, which for a large number of user stations results in a significant reduction of the temporal overheads. This gain can be observed by comparing the curves with and without CSI overheads (curves labeled with NoCSIov) in Figure \ref{Fig:FigureContAlfa02Beta08} for both amendments. The impact of the channel sounding rate in the achievable throughput for an IEEE~802.11ax WLAN is shown in Figure \ref{Fig:FigureCSBIrate}. The reduction of the throughput when $\lambda_{\text{csi}}$ increases is almost independent of the number of sounded stations, which confirms the benefits of using UL MU transmissions to report the CSI information.

Finally, it is worth to mention that using the same CW$_{\min}$ and CW$_{\max}$ parameters for both the AP and user stations result in a higher UL throughput even when only the $20$ \% of DL transmissions are SU (i.e., $\alpha=0.2$), and the $80$ \% of the scheduled MU transmissions are DL (i.e., $\beta=0.8$). Reducing the number of MU transmissions scheduled by the AP (i.e., increasing $\alpha$) also affects negatively the UL throughput when the number of user stations is low (cf. Figure \ref{Fig:FigureAlpha}), as user stations take also advantage of the scheduled UL MU transmissions. When there are many user stations, since the AP is not able to access the channel frequently, the effect of $\alpha$ is low. Moreover, adjusting the fraction of MU transmissions that are DL or UL (i.e., $\beta$), the AP can improve the balance between DL and UL throughput, prioritizing one of them when required, which allows IEEE 802.11ax WLANs to adapt to changing scenarios (Figure \ref{Fig:FigureBeta}). As it can be expected, similarly to $\alpha$, changing the fraction of DL and UL MU transmissions scheduled by the AP has more impact in the WLAN performance when the number of stations is low.

%---------------------------------------------
%---------------------------------------------

\subsection{The Impact of the Maximum A-MPDU Size, Channel Width, and Number of Spatial Streams on Throughput} 

\begin{figure}[t]
\centering
\epsfig{file=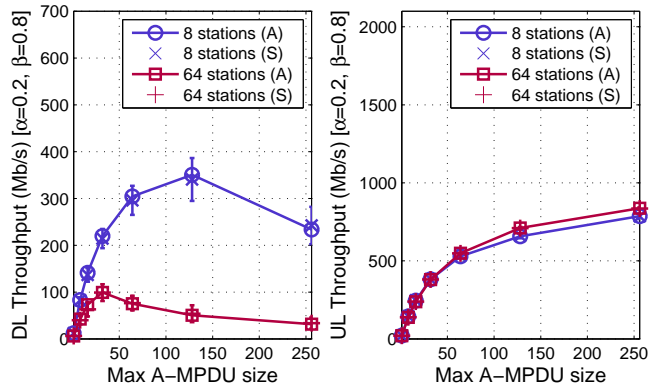,width=0.6\columnwidth,angle=-0}
\caption{Effect of increasing the maximum A-MPDU size on the UL and DL throughput for $B=80$ MHz.}
\label{Fig:FigureMaxAMPDU}
\end{figure}

To further understand the achievable saturation throughput of IEEE 802.11ax WLANs, we study how the maximum A-MPDU size, the channel width and the number of spatial streams impact on it. Figure~\ref{Fig:FigureMaxAMPDU} shows the WLAN throughput when the maximum A-MPDU size increases for $8$ and $64$ user stations, and $B=80$ MHz. While the UL throughput increases monotonically with the maximum A-MPDU size, the DL throughput starts decreasing after a certain maximum A-MPDU size value, which is different for each number of user stations. For example, for $N=64$ user stations the DL throughput reaches its peak for a maximum A-MPDU size of $32$ frames, which increases to $128$ frames for $N=8$ user stations. This result can be explained by considering that a large number of user stations reduces the transmission opportunities of the AP, and therefore, in that situation the AP prefers short transmissions (i.e., with less number of aggregated frames) to avoid further increasing the temporal intervals at which the AP is able to access the channel.

\begin{figure}[t]
\centering
\epsfig{file=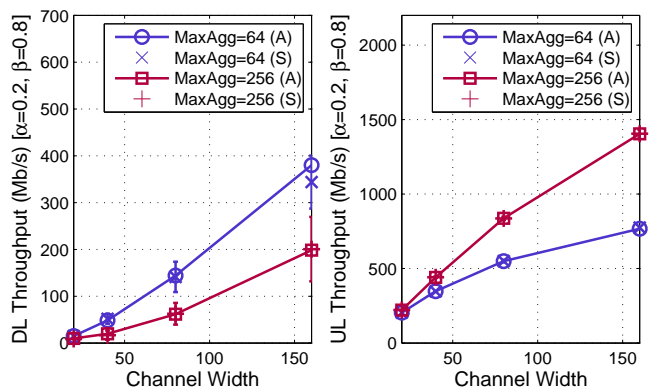,width=0.6\columnwidth,angle=-0}
\caption{Effect of increasing the channel width on the UL and DL throughput for $N=32$ user stations. }
\label{Fig:FigureChannelWidth}
\end{figure}

Considering two maximum A-MPDU sizes of $64$ and $256$ frames, Figure \ref{Fig:FigureChannelWidth} shows the DL and UL throughput when different channel widths are used for $N=32$ user stations. As expected, the DL and UL throughput increase with the channel width. However, similar to the case shown in Figure \ref{Fig:FigureMaxAMPDU}, we can observe again that the DL throughput is lower for higher maximum A-MPDU sizes. Complementing previous justification, this result is explained by the EDCA operation, which gives the same transmission opportunities (in the long-term) to all contenders if they use the same CW$_{\min}$ and CW$_{\max}$ parameters. In addition, since user stations transmit for periods of time proportional to the maximum A-MPDU size, increasing this value reduces even more the amount of effective airtime allocated to the AP. Therefore, allowing large A-MPDU frames can be detrimental for the DL performance, and a cautious approach configuring this parameter is required.

\begin{figure}[t]
\centering
\epsfig{file=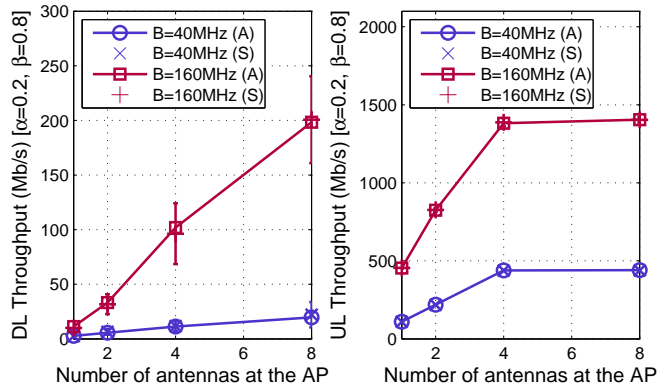,width=0.6\columnwidth,angle=-0}
\caption{Effect of increasing the number of antennas at the AP for $N=32$ stations. The number of antennas at the stations $M_{\text{sta}}$ is kept equal to 4.}
\label{Fig:FigureNumberAntennasAP}
\end{figure}

Adding more antennas at the AP allows for increasing the number of spatial streams used in both SU and MU transmissions. Figure \ref{Fig:FigureNumberAntennasAP} shows that the DL throughput increases with the number of antennas at the AP since $i$) the number of user stations multiplexed in each RU increases proportionally, and $ii$) the number of spatial streams allocated to each user in SU transmissions also increases, though bounded by the number of antennas at the user stations, which is set to $M_{\text{sta}}=4$. Moreover, it can also be observed that the gain of adding more antennas increases proportionally to the channel width.

%---------------------------------------------
%---------------------------------------------

\subsection{UL Throughput Improvement}

\begin{figure}[t]
\centering
\subfigure[$\alpha=0.2$, and $\beta=0.8$]{\epsfig{file=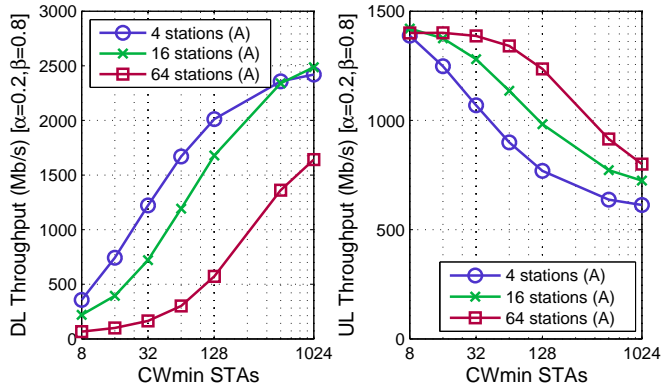,width=0.6\columnwidth,angle=-0}\label{Fig:FigureCWeffects1}}\\
%
%\hspace{2mm}
%
\subfigure[$\alpha=0.2$, and $\beta=0.2$]{\epsfig{file=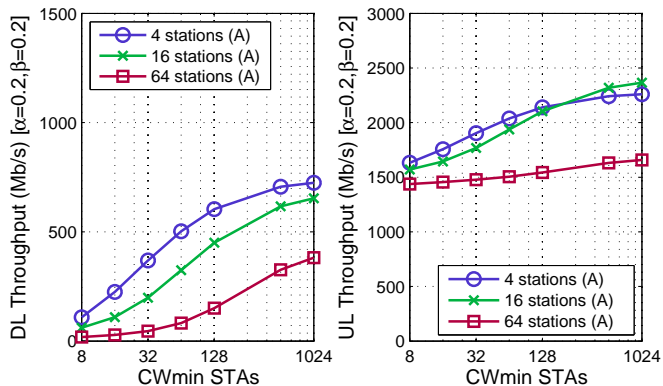,width=0.6\columnwidth,angle=-0}\label{Fig:FigureCWeffects2}}
\caption{Effect of increasing the station's $\text{CW}_{\min}$ on the DL and UL throughput.}
\label{Fig:CWSTAsIncrease}
\end{figure}

Collisions between transmissions initiated by the AP and those from user stations severely harm the WLAN performance. Of special significance is the case when a trigger packet sent by the AP to initiate an UL MU transmission collides with a packet transmitted by one of the stations scheduled by the AP and included in the trigger frame. In order to avoid such a situation, a mechanism is required to reduce the chances that user station transmissions collide with AP transmissions. To do so, we propose to increase the $\text{CW}_{\min}$ for the user stations supporting UL MU transmissions\footnote{User stations not supporting UL MU transmissions should keep the recommended $\text{CW}_{\min}$ size to fairly access the channel.}. The benefits of such an approach are observed in Figures \ref{Fig:FigureCWeffects1} and \ref{Fig:FigureCWeffects2}, where reducing the rate at which user stations attempt transmissions results in a higher uplink throughput. Only when the AP is prioritizing downlink MU transmissions (i.e., the $\beta$ parameter is close to $1$), such a policy is not beneficial for user stations. Finally, it is worth to mention that in those cases in which the UL throughput is higher when the user station's $\text{CW}_{\min}$ is increased, the expected packet transmission delay will be lower, since the user stations are scheduled by the AP more often than when they are aggressively contending to access the channel.

%---------------------------------------------
%---------------------------------------------
%---------------------------------------------
%---------------------------------------------

\section{Conclusions}
\label{Sec:Conclusions}

This paper overviews the new PHY/MAC characteristics of IEEE 802.11ax WLANs, with emphasis on their MU transmission capabilities. Based on them, we introduce a possible operation of IEEE 802.11ax WLANs that exploits efficiently both MU-MIMO and OFDMA techniques. The presented results provide novel insights on how IEEE 802.11ax WLANs perform for different configurations, including aspects such as the effect of the channel sounding rate, and the relationship between the maximum A-MPDU size, channel width and number of antennas at the AP. We also show the need to provide strict prioritization to the AP to avoid collisions with user stations when the AP is in charge of scheduling MU transmissions for both DL and UL to optimize the WLAN throughput.

Future work on the performance evaluation of MU transmissions in IEEE 802.11ax WLANs should focus on the analysis of the effects of miscellaneous traffic (i.e., non-saturation case) and channel conditions, including multiple traffic classes and different available transmission rates per station. In such a scenario, the design of efficient schedulers to select the stations that will be included in a MU transmission are key for maximizing the WLAN performance and reach IEEE 802.11ax expectations. In addition, special emphasis on the analysis of the advantages and disadvantages of MU transmissions in high density WLAN scenarios should be placed, accounting also for coexistence issues, and potential spatial reuse improvements. Moreover, novel analytical and simulation tools may have to be developed to study all those aspects, as current ones many not be able to cope with such complex scenarios.

The IEEE 802.11ax amendment is still in its development phase. Although most of its fundamental characteristics are already included and consolidated in the current version of the IEEE 802.11ax draft amendment, there are still many aspects open that need to be refined and detailed in the next few years, specially those related about how to efficiently use the new functionalities provided. We expect that this work will contribute to a better understanding of the performance of future IEEE 802.11ax WLANs.

%---------------------------------------------
%---------------------------------------------
%---------------------------------------------

\section{Acknowledgements}

The work of K. Kosek-Szott has been carried out as part of a project financed by the Polish National Science Centre (decision no. DEC-2011/01/D/ST7/05166), and the work of B. Bellalta has been supported by the Catalan Government SGR grant for research support (2017-SGR-1188), and by a Gift from the Cisco University Research Program (CG\#890107, Towards Deterministic Channel Access in High-Density WLANs) Fund, a corporate advised fund of Silicon Valley Community Foundation.

\appendix
\section{Abbreviations}
In Table \ref{tab:abbreviations} we provide the full names of the abbreviations used in the paper.

\begin{table}[htb]
  \centering
  \caption{Abbreviations used in the paper}
    \begin{tabular}{ll}
\hline
    \textbf{Abbreviation} & \textbf{Full name}
      \\
\hline
    AC &       Access category
      \\
    ACK &       Acknowledgment
      \\
    AIFS &      Arbitrary inter-frame space
      \\
    A-MPDU &      Aggregated MDPU
      \\
    AP &      Access point
      \\
    BACK &      Block ACK
      \\
	BSS & Basic service set
	\\
    BSR &      Buffer status report
      \\
    CSI &      Channel state information
      \\
    CSIREP &      CSI reply
      \\
    CTS &      Clear to send
      \\
    CW &      Contention window
      \\
    DCF &      Distributed coordination function
      \\
    DCM &      Dual carrier modulation
      \\
    DL &      Downlink
      \\
    EDCA &      Enhanced distributed channel access
      \\
    EDCAF &      EDCA function
      \\
    GI &      Guard interval
      \\
    LDPC &      Low density parity check
      \\
    MAC &      Medium access control
      \\
    MCS &      Modulation and coding scheme
      \\
    MIMO &      Multiple input, multiple output
      \\
    MPDU &      MAC protocol data unit, MAC frame
      \\
    MU &      Multiuser
      \\
    MS-BACK &      Multiuser block ACK
      \\
    MU-MIMO &      Multiuser multiple input, multiple output
      \\
    MU-RTS &      Multiuser RTS
      \\
    NDP &      Null data packet
      \\
    NDPA &      Null data packet announcement
      \\
    OBSS &      Overlapping basic service set
      \\
    OFDM &      Orthogonal frequency division multiple access
      \\
    OFDMA &      Orthogonal frequency division multiple access
      \\
    PHY &      Physical layer
      \\
    PPDU &      Physical layer protocol data unit
      \\
    QoS &      Quality of service
      \\
    RTS &      Request to send
      \\
    RU &      Resource unit
      \\
    SIFS &      Short inter-frame space
      \\
    SS &      Spatial stream
      \\
    SU &      Single-user
      \\
    TG &      Task group
      \\
    TGax &      IEEE 802.11ax Task Group
      \\
    UL &      Uplink
      \\
    UL MU-MIMO &      Uplink MU-MIMO
      \\
    WLAN &      Wireless local area network
      \\
\hline
    \end{tabular}%
  \label{tab:abbreviations}%
\end{table}%

\bibliographystyle{unsrt}
\bibliography{Bib}

\end{document}